\documentclass[10pt,letterpaper]{article}
\usepackage{amsmath,amssymb}
\usepackage{changepage}
\usepackage[utf8x]{inputenc}
\usepackage{textcomp,marvosym}
\usepackage{cite}
\usepackage{nameref,hyperref}
\usepackage{csquotes}
\usepackage[right]{lineno}
\usepackage{microtype}
\DisableLigatures[f]{encoding = *, family = * }
\usepackage[table]{xcolor}
\usepackage{array}
\usepackage[aboveskip=1pt,labelfont=bf,labelsep=period,justification=raggedright,singlelinecheck=off]{caption}

\usepackage{graphicx}
\providecommand{\keywords}[1]{\textbf{\textit{Keywords}} #1}
\begin{document}
\begin{flushleft}
{\Large
\textbf\newline{Dynamics of racial segregation and gentrification in New York City}
}
\newline
\\
Felipe G. Operti$^1$, Andr\'e A. Moreira$^1$, Andrea Gabrielli$^2$, Hernan Makse$^3$, and Jos\'e S. Andrade Jr.$^1$
\\
\bigskip
\textbf{1} Departamento de F\'isica, Campus do Pici, Universidade Federal do Cear\'a, 60451-970, Fortaleza, Cear\'a, Brazil
\\
\textbf{2} Istituto dei Sistemi Complessi (ISC) - CNR, UoS Sapienza, Dipartimento di Fisica, Università Sapienza, P.le Aldo Moro 5, 00185, Rome, Italy
\\
\textbf{3} Levich Institute and Physics Department, City College of New York, 10031, New York, New York, USA
\\
\bigskip
\Yinyang These authors contributed equally to this work.

\end{flushleft}
\section*{Abstract}

Racial residential segregation is interconnected with several other phenomena such as income inequalities, property values inequalities, and racial disparities in health and in education. Furthermore, recent literature suggests the phenomena of gentrification as a cause of perpetuation or increase of racial residential segregation in some American cities. In this paper, we analyze the dynamics of racial residential segregation for white, black, Asian, and Hispanic citizens in New York City in the years of 1990, 2000, and 2010. It was possible to observe that segregation between white and Hispanic citizens, and discrimination between white and Asian ones has grown, while segregation between white and black is quite stable. Furthermore, we analyzed the per capita income and the Gini coefficient in each segregated zone, showing that the highest inequalities occur in the zones where there is overlap of high-density zones of pair of races. Focusing on census tracts that have changed density of population during these twenty years, and, particularly, by analyzing white and black people's segregation, our analysis reveals that a positive flux of white (black) people is associated to a substantial increase (decrease) of the property values, as compared with the city mean. Furthermore, by clustering the region of high density of black citizens, we measured the variation of area and displacement of the four biggest clusters in the period from 1990 to 2010. The large displacements ($\approx 1.6$ $km$) observed for two of these clusters, namely, one in the neighborhood of Harlem and the other inside the borough of Brooklyn, led to the emergence of typically gentrified regions.

\keywords{Racial residential segregation; Gentrification; City Clustering Algorithm}

\section*{Introduction}

Although it is not a recent phenomenon, racial residential segregation (RRS) continues to permeate the United States metropolitan areas and it is still an object of study for scientists of different areas \cite{chodrow2017,roberto2017,roberto2016,louf2016,roberto2015,firebaugh2015,lichter2015,fowler2014,boustan2012,
logan2010,watson2009,readon2008a,readon2008b,jargowsky2005,logan2004,readon2004,acevedo-garcia2003,charles2003,readon2002,williams2001,fang1998,massey1993,massey1988,winship1977,duncan1955}. The decrease of RRS in American cities is controversial and drastically varies from one city to another. Furthermore, it shows different trends according to the race analyzed. For example, several studies show that the segregation between white and black citizens has decreased in the last fifty years \cite{roberto2017,firebaugh2015,logan2010,logan2004}. Instead, segregation between white and Hispanic, and white and Asian citizens has increased \cite{logan2010,logan2004}.  

Several indexes were developed to quantify RRS \cite{chodrow2017,roberto2016,louf2016,roberto2015,jargowsky2005,readon2004,readon2002,massey1988,winship1977,duncan1955}. The first and still most used nowadays is the dissimilarity index created by Duncan and Duncan in 1955 \cite{duncan1955}. Subsequently, in 1988, Massey and Denton \cite{massey1988} defined five distinct axes of measurement of residential segregation: evenness, exposure, concentration, centralization, and clustering. The authors affirmed that, in order to fully analyze residential segregation, at least five indexes corresponding to the five spatial dimensions are necessary. Meanwhile, in 2004, Reardon and O’Sullivan’s developed several measures of multigroup segregation and, among them, the authors consider the Information Theory Index the most conceptually and mathematically satisfactory measure to quantify residential segregation \cite{readon2004}.

RRS is the cause and effect of several inequalities. Studies show the relations between racial segregation and income inequalities \cite{watson2009} and property values inequalities. Furthermore, RRS causes racial disparities in health and in education \cite{watson2009,acevedo-garcia2003,williams2001,fang1998}. In New York City, for instance the mortality rates of black citizens vary substantially by locality according to the pattern of racial segregation \cite{fang1998}.

In the recent years, some researches also suggest that the phenomena of gentrification is a cause of perpetuation or even of the increase of RRS \cite{helbrecht2018,brown2017,lees2016,freeman2008}. Gentrification is defined by \textit{The Encyclopedia of Housing} \cite{freeman2004,smith1998} as:
\begin{displayquote}
\textit{The process by which central urban neighborhoods that have undergone disinvestment and economic decline experience a reversal, reinvestment, and the in-migration of a relatively well-off, middle and upper middle-class population.}
\end{displayquote}
The main reason to indicate gentrification as a cause of perpetuation of racial segregation is the presumed displacement of the low-income class, in many cases predominantly black or Hispanic citizens, from their native neighborhood during the gentrification process \cite{helbrecht2018,freeman2008,freeman2004,curran2006,zuk2017}. Taking the example of New York City once again, there is an intense debate about the gentrification of regions inside the neighborhoods of Harlem and the borough of Brooklyn \cite{zukin2009,lees2003,schaffer1986}.

The aim of this paper is to study the dynamics of RRS in New York City from 1990 to 2010. Here, we developed a novel method able both to measure RRS and to delimit the segregated zones. Indeed, differently from previous measures, our method, in addition to quantifying the phenomena, provides a topography of the segregation. Furthermore, in the section \textit{Comparison with the Dissimilarity index}, we compare our segregation index, the Overlap coefficient, with the dissimilarity index.

With the limit of the segregated zones, we analyze the per capita income in each high-density zone of population (defined for each race) and also in the zones of overlaps between them. In order to quantify income inequality, we calculate the Gini coefficient in each zone. Then, we study the variation of the per capita income and of the properties' value for the census tracts that change zone during these twenty years. Finally, we focus on the segregation between white and black citizens. Particularly, we use a simplified version of the City Clustering Algorithm (CCA) \cite{operti2017,caminha2017,oliveira2014,rozenfeld2011,rozenfeld2008,makse1998,makse1995,arcaute2015,arcaute2016} to cluster the high-density zone of black citizens and to measure the displacement and the area of the four biggest clusters (one of these clusters includes the neighborhood of Harlem and another one is inside the borough of Brooklyn).

The paper is structured as follows: first, we introduce our method. Then, we present the results of the application of the method to New York City. Finally, we draw the conclusion about the results. In the Appendix A we provide the information for the acquisition of the data.

\section*{Method}

The method consists of the following steps: first we define the limits of the city using the City Clustering Algorithm (CCA) \cite{operti2017,caminha2017,oliveira2014,rozenfeld2011,rozenfeld2008,makse1998,makse1995,arcaute2015,arcaute2016}. Second we find the high-density zones for white, black, Asian, and Hispanic citizens. Finally, we measure the RRS through the Overlap Coefficient.

The CCA is an algorithm introduced to define boundaries of metropolitan areas \cite{operti2017,caminha2017,oliveira2014,rozenfeld2011,rozenfeld2008,makse1998,makse1995,arcaute2015,arcaute2016}. Its result depends on two parameters: a population density threshold $D^*$ (in $people/km^2$), and a cutoff length $\ell$ (in $km$). The elementary information for population data are provided in \textit{census tract}. Where the tracts are geographic regions defined by the United States Census Bureau \cite{census} (see \textit{Appendix A} for more information about the database). For each tract, we have the total area and the total population given by the sum of people of each race. Therefore, for each tract, its population density is calculated. According to the CCA, the assumption is that only the tracts with $D_i>D^*$ are populated. 

The next step of the algorithm is the clusterization. In this step, we define the urban center. For each populated tract, we draw a circle of radius $\ell$ with center in the centroid of the tract. All populated tracts that have the centroid inside the circle belong to the same cluster, and, therefore, the same city. The parameter $D^*$ and $\ell$ are chosen respecting the isometry between area and population of the cities \cite{operti2017,caminha2017,oliveira2014}. The algorithm is applied in the entire country and, subsequently, we extract only the cluster equivalent to New York City. 

The importance of using the CCA to define the urban area of New York City is due to the fact that RRS deeply depends on the definition of urban areas \cite{roberto2017,roberto2016,roberto2015}. For example, it was shown in \cite{operti2017,oliveira2014} that the Metropolitan Urban Areas (MSA) have large inhabited regions. Instead, the aim of our research is to analyze RRS in a very dense urban area, specifically in New York City.

We define the high-density (HD) zones as regions inside the city with a high population density of a specific race. The HD zone of a specific race $r$ is defined applying a density threshold $D_r^*$. We consider the tracts with $D_r>D_r^*$ populated of that race. $D_r$ is the population density of that race. The choice of parameter $D_r^*$ is made by studying how the fraction of population of race $r$, with respect of the total population of the same race inside the whole city, depends on it. Therefore, for each race $X$, we define a parameter $p_r$ as:
\begin{equation}
p_r=\frac{Population~of~race~r~inside~the~HD~r~zone}{Total~population~of~race~r~inside~the~city}.
\label{eq_p}
\end{equation} 
To make the analysis as uniform as possible, we choose $D^*_r$ so that both $D^*_r$ and $p_r$ take similar values for all considered races $r$.

\begin{figure}[!h]
\centerline{\includegraphics[width=0.8\textwidth]{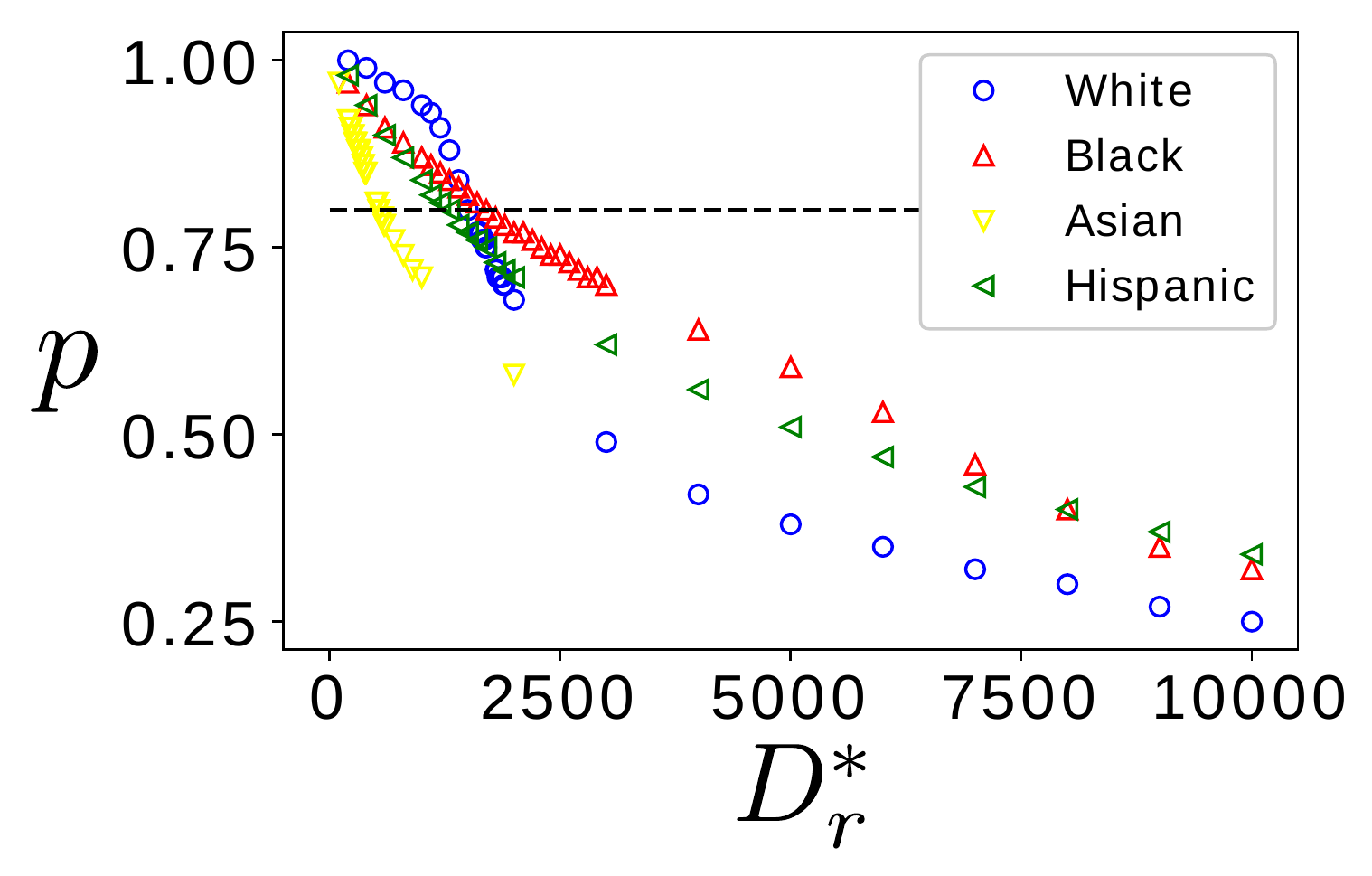}}
\caption{{\bf Variation of $p$ in function of parameter $D^*_r$ for each race in New York City in 2010.} The Figure shows the variation of parameter $p$ in function of parameter $D^*$ for white, black, Asian, and Hispanic. The dashed black line in $p=0.8$ shows the $80\%$ of the total population for each race.}
\label{parameter}
\end{figure}

In Fig \ref{parameter} we show the variation of the parameter $p$ in function of parameter $D^*_r$ for each race in New York City. We consider the same fraction of people in three cases using a similar $D^*_r$: when it is next to $0$, to $\infty$, and $\sim 2000$. The first two are trivial, in fact they show respectively all and any population. While in $p=0.8$ ($80\%$ of the total population for each race), for each race $D^*\sim 2000$. The dotted black line in the Figure is exactly in $p=0.8$ showing the $80\%$ of the total population of each race.
Parameter $p_r$ has been tested in the interval from $0.7$ to $0.9$ without find deep discrepancies in the results. Therefore, at the end of this step, the method provides well-defined geographic limits of the HD zones for each race. 

From the definition of the HD zones, we measure the RRS between two races computing the sharing area (or overlap area) between the two HD zones. Therefore, we define the Overlap coefficient (or Szymkiewicz-Simpson coefficient \cite{vijaymeena2016}) as:
\begin{equation}
O_{rr'}=\frac{\left | X_r \cap  X_{r'} \right |}{min(\left | X_r \right |,\left | X_{r'} \right |)},
\label{eq_overlap1}
\end{equation}
where $X_r$ and $X_{r'}$ are respectively the HD zone areas of races $X_r$ and $X_{r'}$. Coefficient $O_{rr'}$ is the sharing area between the HD $r$ zone and the HD $r'$ zone divided by minimum area between the two zones. The Overlap coefficient is included between $0$ and $1$. When it is next to $0$ (low overlap), the coefficient indicates high segregation, while when it is next to $1$ (high overlap), it indicates low segregation (see Table \ref{overlapCoeff}).

\section*{Results}

\begin{figure}[!h]
\centerline{\includegraphics[width=1\textwidth]{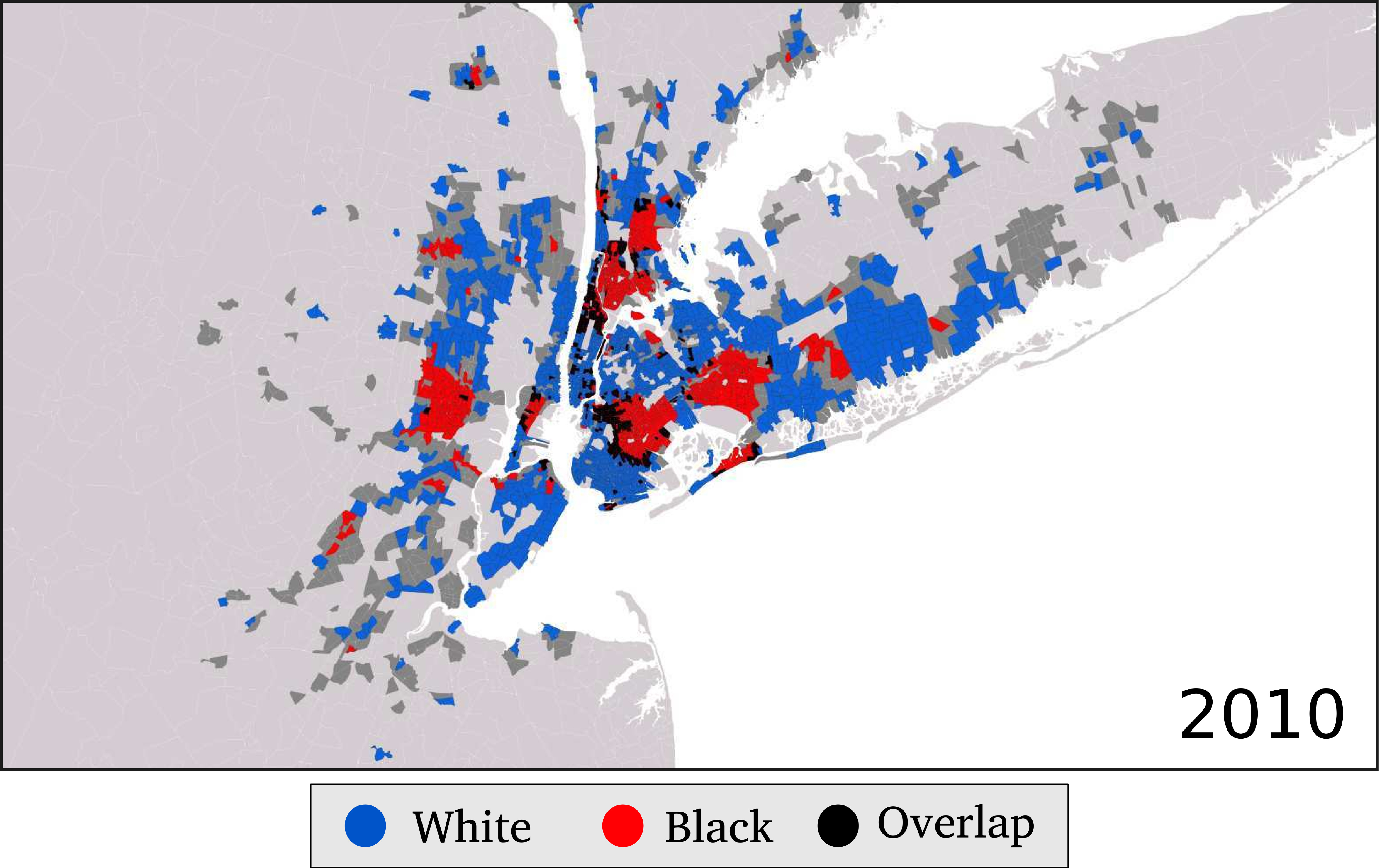}}
\caption{{\bf HD zone for white and black.} We show the HD zone for white (blue) and black (red) citizens with the respective Overlap zone (black) in the year 2010. Dark grey tracts are part of the city that do not belong to any of the zones, while light grey tracts are not part of New York City.}
\label{dynamics}
\end{figure}

Firstly, we define the limits of New York City by applying the CCA to the population data in 2010 (see \textit{Appendix A} for more details about the data). Then, we calculate the HD zone for white, black, Asian, and Hispanic for the year of 1990, 2000, and 2010. In Fig \ref{dynamics}, we show the HD zone for white and black citizens with the respective Overlap zone in the year 2010. 

For each pair of races, we calculate the Overlap coefficients and the results were presented in Table \ref{overlapCoeff}.
\begin{table}[]
\centering
\caption{Overlap Coefficients}
\label{overlapCoeff}
\begin{tabular}{l|l|l|l}
                   & \textbf{1990} & \textbf{2000} & \textbf{2010} \\ \hline
\textbf{White and Black}    & 0.22 & 0.19 & 0.20 \\ \hline
\textbf{White and Hispanic} & 0.61 & 0.53 & 0.47 \\ \hline
\textbf{White and Asian}    & 0.82 & 0.73 & 0.67 \\ \hline
\textbf{Black and Hispanic} & 0.52 & 0.52 & 0.61 \\ \hline
\textbf{Black and Asian}    & 0.27 & 0.24 & 0.26 \\ \hline
\textbf{Hispanic  and Asian} & 0.58 & 0.48 & 0.29 \\ \hline
\end{tabular}
\end{table}
The Table shows that the segregation between white and black, and black and Asian citizens remains quite stable during the time interval. While segregation between white and Hispanic, white and Asian, and Hispanic and Asian has increased, the segregation between black and Hispanic citizens has decreased. Black people are constantly the most segregated having a high overlap coefficient only with Hispanic.

\begin{figure}[!h]
\centerline{\includegraphics[width=1\textwidth]{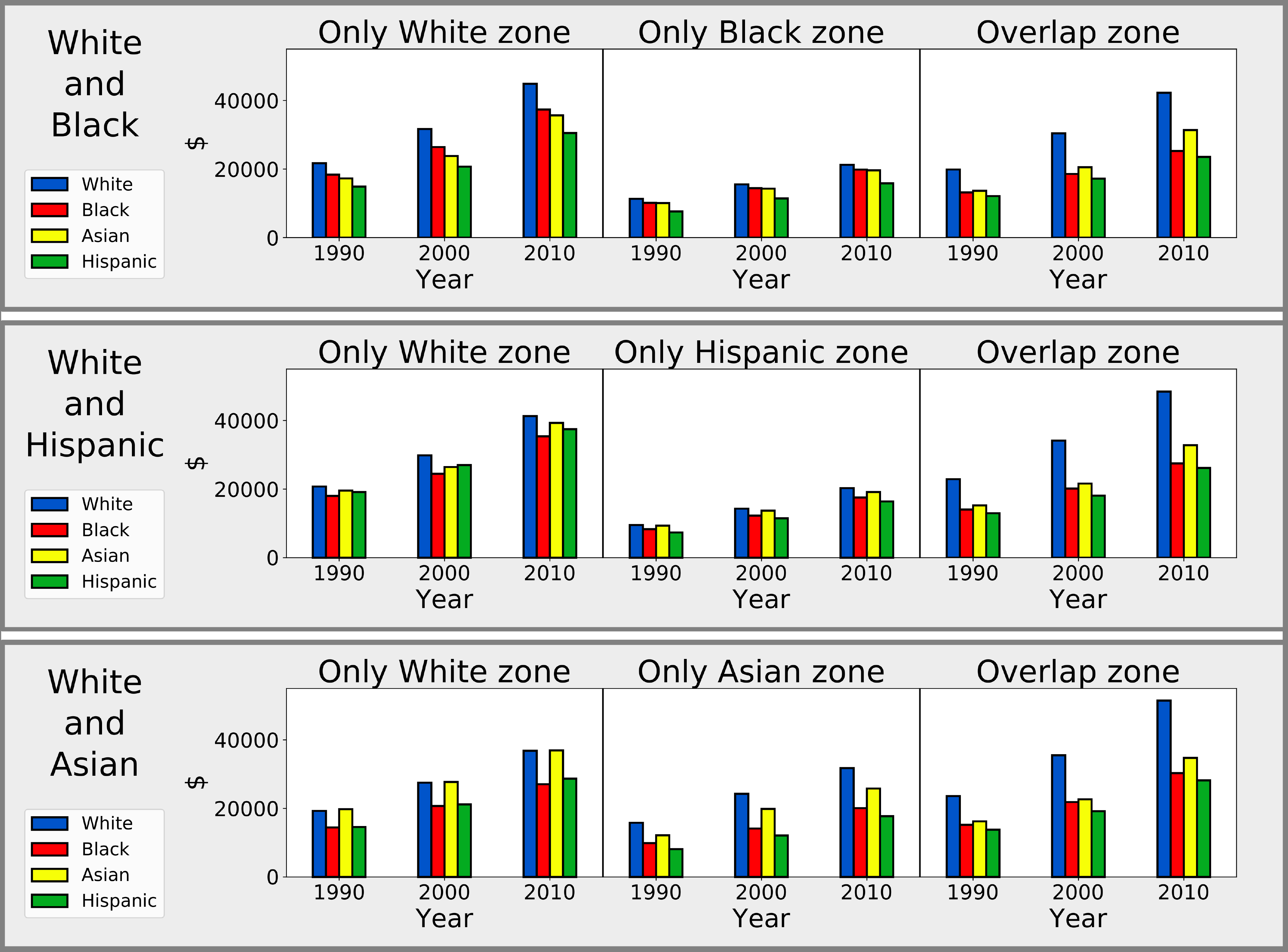}}
\caption{{\bf Per capita income analysis.} The Figure shows the mean per capita income for each race for the study of the segregation between white and black, white and Hispanic, and white and Asian for the years of 1990, 2000, and 2010.}
\label{total_income}
\end{figure}

After the definition of the HD zones and the Overlap zones, we calculate the average per capita income of each race inside each zone for the years of 1990, 2000, and 2010. The results are presented in Fig \ref{total_income}, where ``only'' means the HD zone without the Overlap zone. The Figure shows that white citizens earn more than all the other races in all the zones except in the study of the segregation between white and Asian citizens. Black and Hispanic citizens earn less than whites in all the zones. Moreover, the Figure shows that income inequality between white and black citizens is greater in the Overlap zone than in the only white zone and the only black zone.

\begin{figure}[!h]
\centerline{\includegraphics[width=1\textwidth]{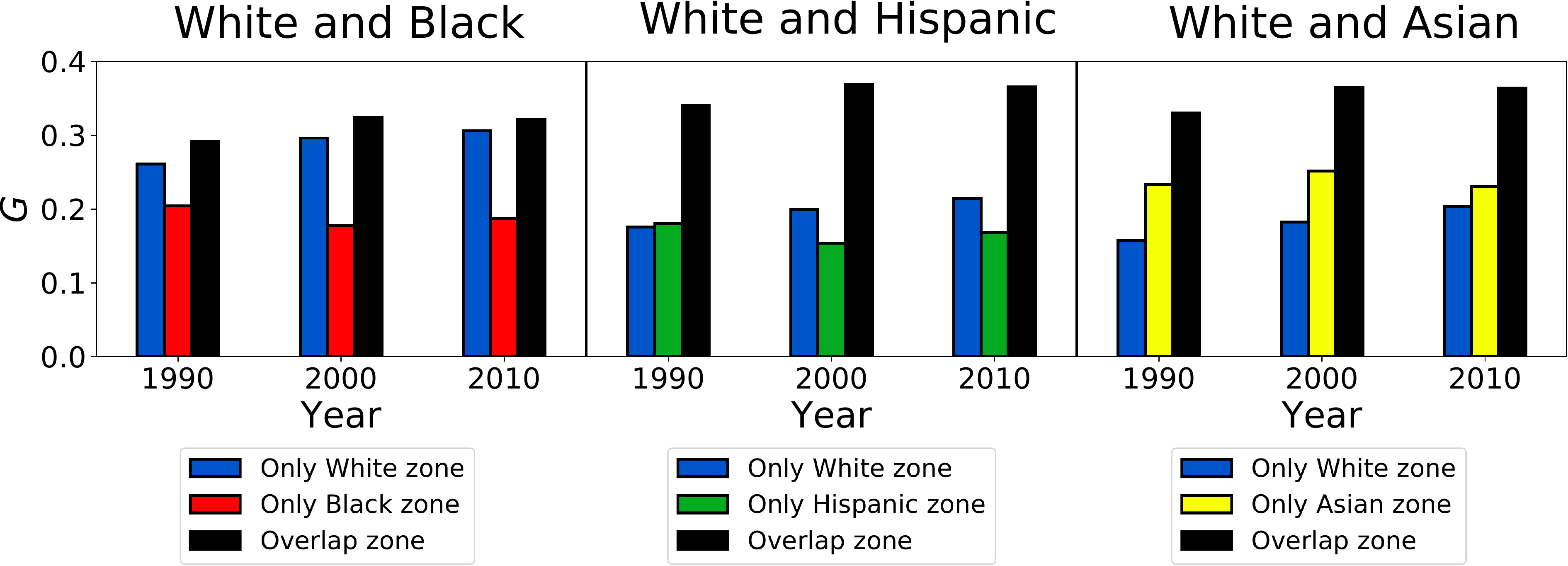}}
\caption{{\bf Gini coefficient for the years of 1990, 2000, and 2010.} The Figure shows the Gini coefficient in the HD only zones and in the Overlap zones for the study of segregation between: white and black, white and Hispanic, and white and Asian.}
\label{gini}
\end{figure}

To study the per capita income inequalities for each study of segregation (white and black, white and Hispanic, and white and Asian), we calculate the Gini coefficient \cite{gini1912} inside each of them. The results are presented in Fig \ref{gini}. The Gini coefficient varies from $0$ to $1$. When it is next to $0$, there is not inequality, while when it is next to $1$, inequality is maximum \cite{gini1912}. The Figure shows that inequality is greater in the Overlap zones in all cases in favor of whites.

\begin{figure}[!h]
\centerline{\includegraphics[width=1\textwidth]{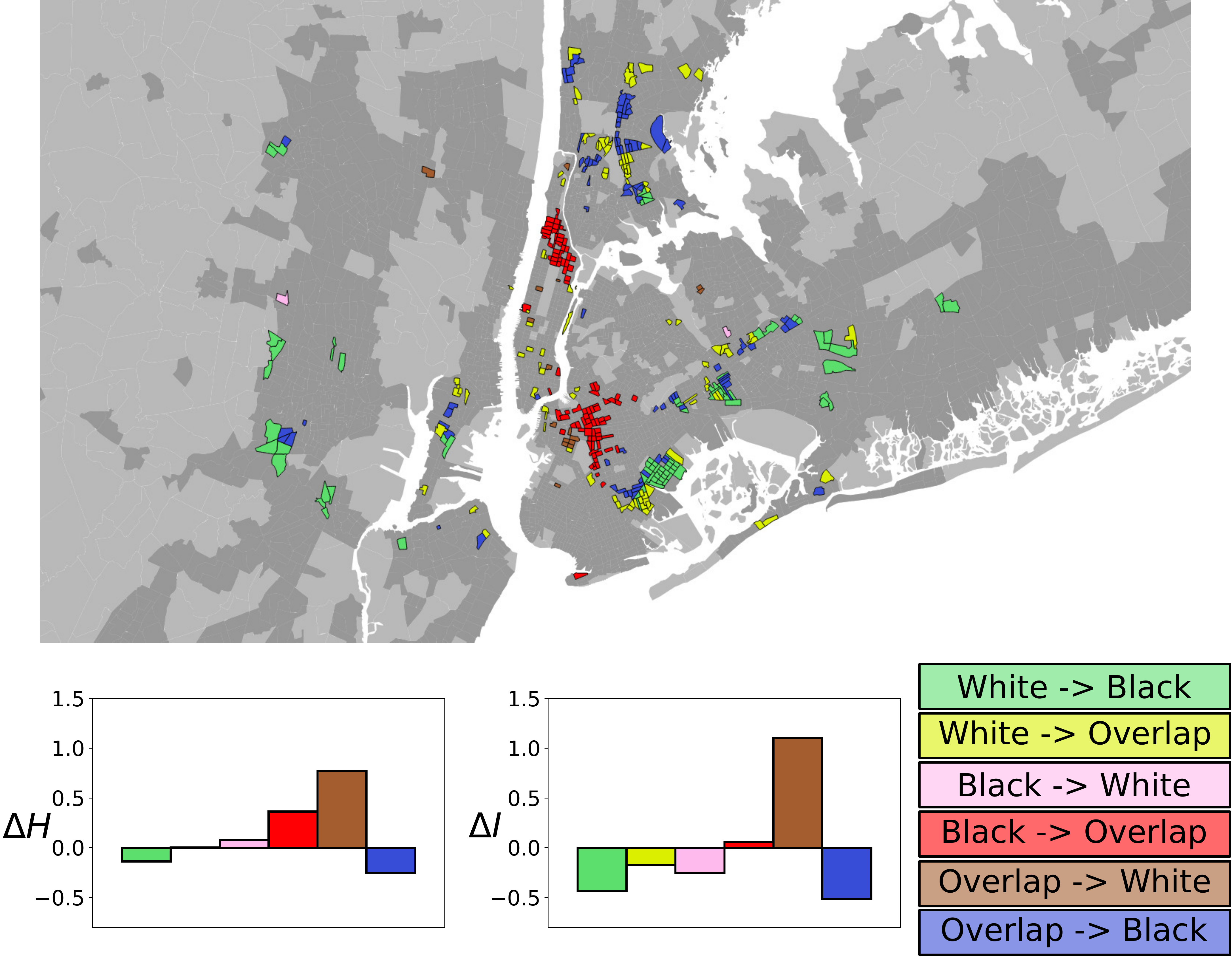}}
\caption{\textbf{Tracts that migrated from one zone to another or to the Overlap zone from 1990 to 2010: white and black citizens.} All the tracts that changed zone during the period from 1990 to 2010 are shown on the map, while the colors show the different alternatives of migration. Furthermore, for each alternative of migration, the value of $\Delta H$ and $\Delta I$ is shown.}
\label{white_black_variation}
\end{figure}

\begin{figure}[!h]
\centerline{\includegraphics[width=1\textwidth]{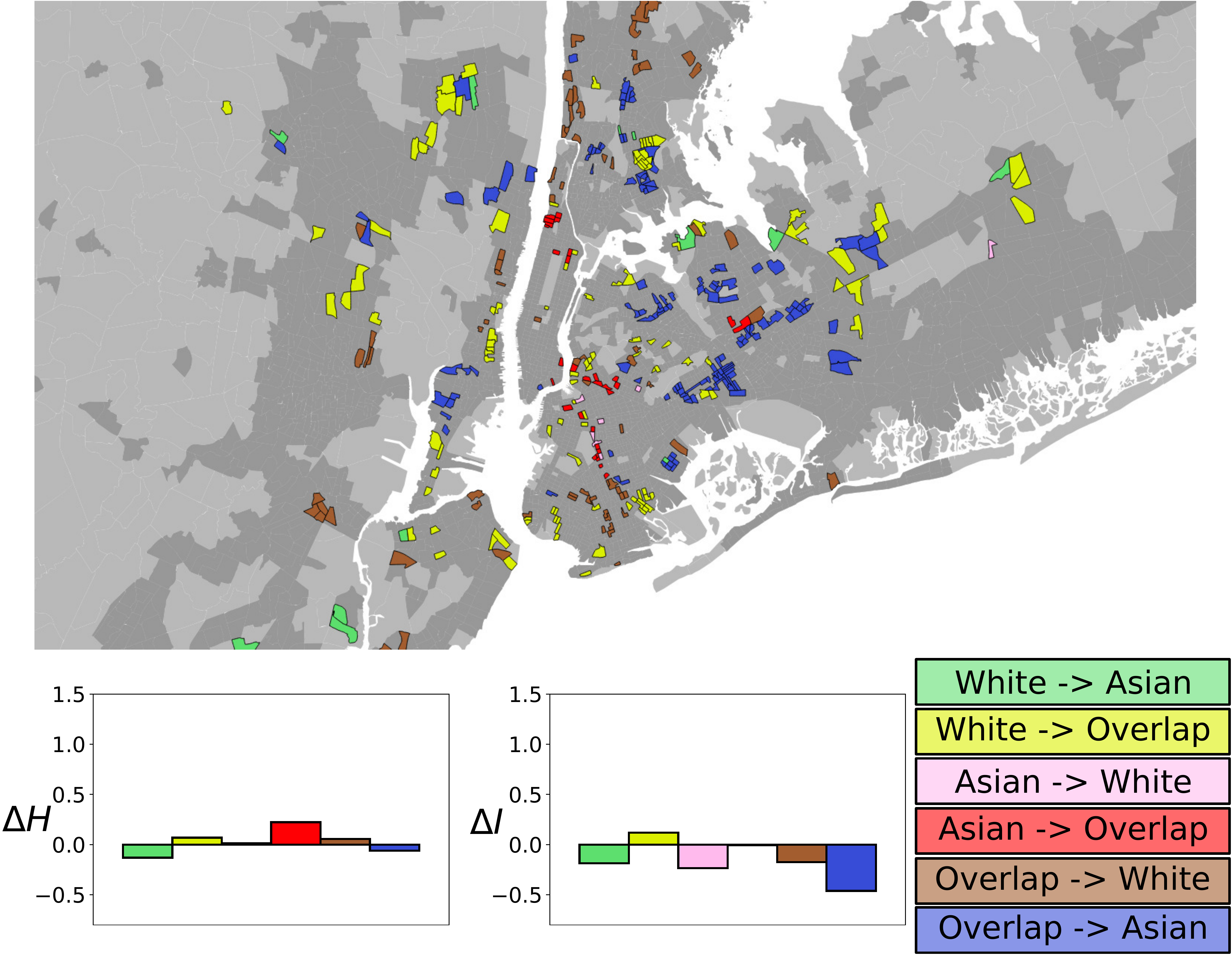}}
\caption{\textbf{Tracts that migrated from one zone to another or to the Overlap zone from 1990 to 2010: white and Asian citizens.} Similar to Fig \ref{white_black_variation}, here we analyze white and Asian citizens. }
\label{white_asian_variation}
\end{figure}

\begin{figure}[!h]
\centerline{\includegraphics[width=1\textwidth]{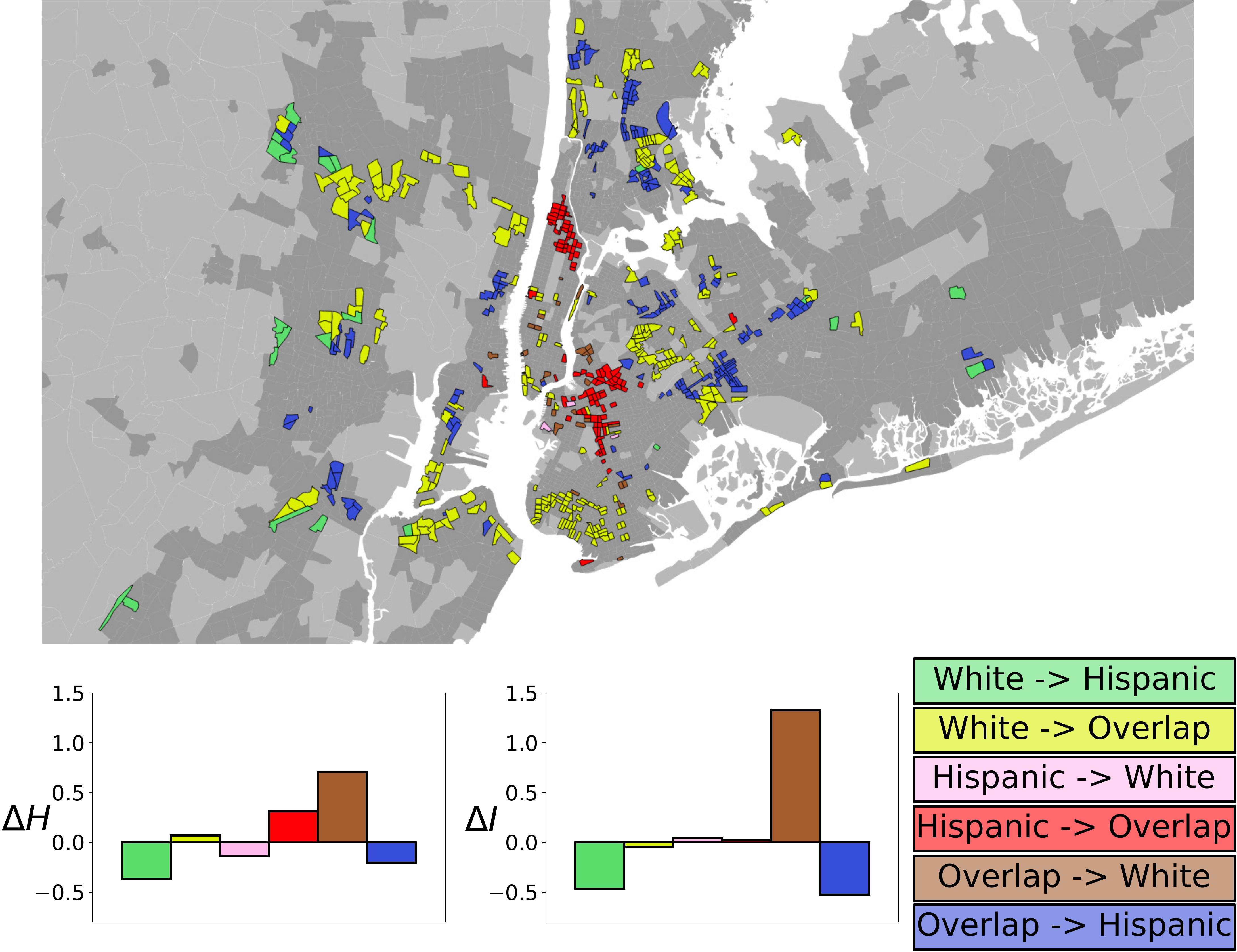}}
\caption{\textbf{Tracts that migrated from one zone to another or to the Overlap zone from 1990 to 2010: white and Hispanic citizens.} Similar to Fig \ref{white_black_variation} and \ref{white_asian_variation}, here we analyze white and Hispanic citizens.}
\label{white_hispanic_variation}
\end{figure}

Furthermore, we analyze the tracts that migrated from one zone to another from 1990 to 2010 for the studies of segregation between: white and black citizens in Fig \ref{white_black_variation}; white and Asian citizens in Fig \ref{white_asian_variation}; and white and Hispanic citizens in Fig \ref{white_hispanic_variation}. The colors in the maps in Figs \ref{white_black_variation}-\ref{white_asian_variation}-\ref{white_hispanic_variation} show the alternatives of migration of the tracts from one zone to another, which are described in the caption. For each alternative, we calculate the average variation of the per capita income ($\Delta I$) and the average variation of the properties values ($\Delta H$) normalized by the average variation in the city ($\overline{\delta I}$ and $\overline{\delta H}$) from 1990 to 2010. The variations are defined as:
\begin{equation}
\Delta I = \frac{1}{N}\sum_{i=1}^{N}\frac{\delta I_i - \overline{\delta I}}{\left | \overline{ \delta I} \right |},
\label{deltaI}
\end{equation} 
and,
\begin{equation}
\Delta H = \frac{1}{N}\sum_{i=1}^{N}\frac{\delta H_i - \overline{\delta H}}{\left | \overline{ \delta H} \right |}.
\label{deltaH}
\end{equation}
Where $N$ is the number of tracts of the analyzed pairs of races and $\delta I_i$ and $\delta H_i$ are the variations of the per capita income and properties values of tract $i$, respectively. Therefore, positive $\Delta I$ or $\Delta H$ mean growth higher than the city mean, while, conversely negative $\Delta I$ or $\Delta H$ mean growth lower than the city mean.

Moreover, we focus on the segregation between white and black citizens and the flux of people from 1990 to 2010 inside the tracts that migrated from one zone to another or to the Overlap zone. The flux of people of a specific race inside a tract is the variation of people of that specific race $X$ inside tract $i$ compared with the mean variation of that specific race in the whole city. Similarly to Eq \ref{deltaI} and \ref{deltaH}, the average flux $\Delta Flux_X$ is defined:
\begin{equation}
\Delta Flux_X = \frac{1}{N}\sum_{i=1}^{N}\frac{\delta Flux_{X,i} - \overline{\delta Flux_X}}{\left | \overline{ \delta Flux_X} \right |},
\label{deltaFlux}
\end{equation} 
where $\overline{\delta Flux_X}$ is the mean flux of race $X$ in the whole city. 

\begin{figure}[!h]
\centerline{\includegraphics[width=1\textwidth]{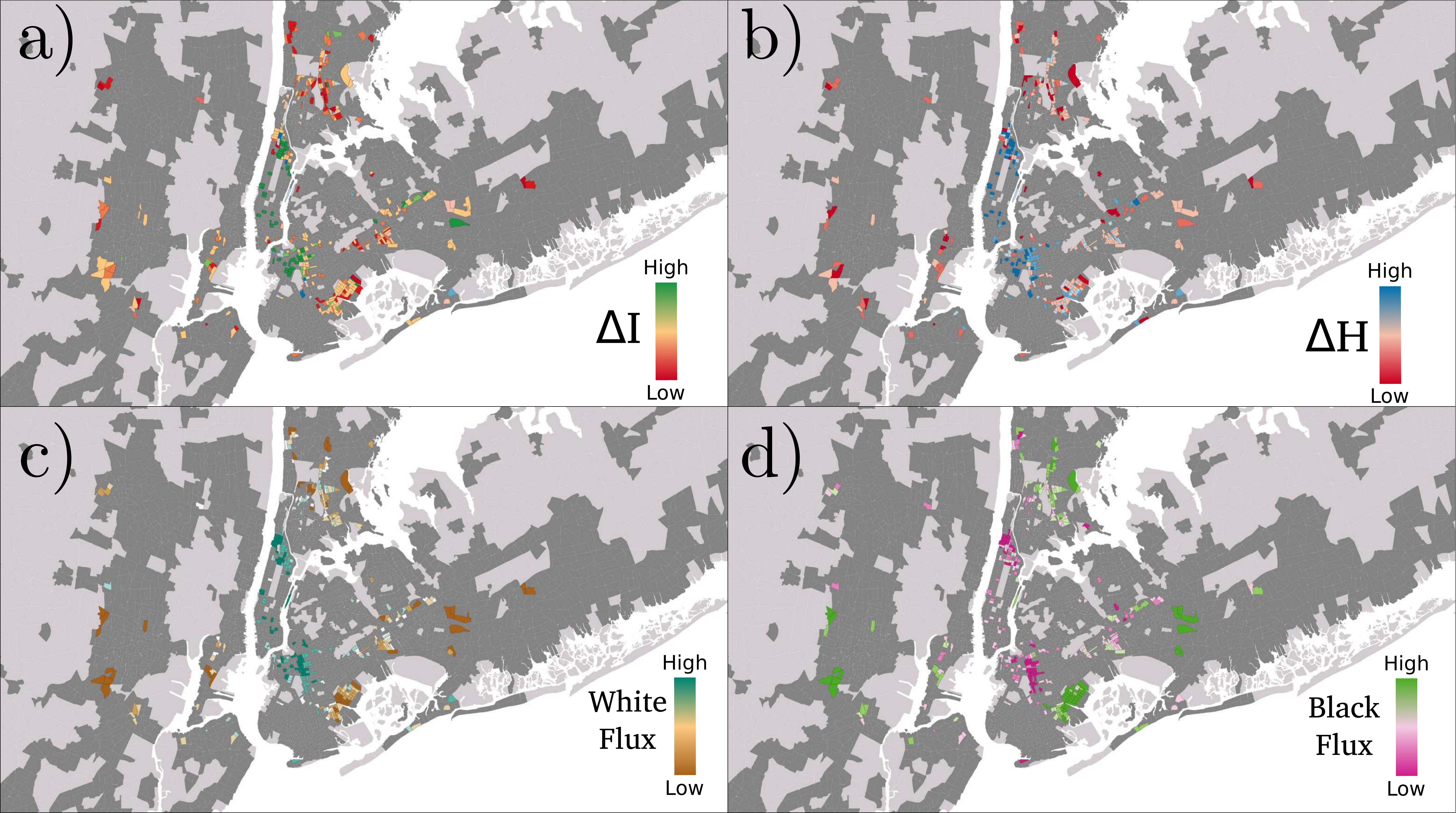}}
\caption{\textbf{Segregation between white and black.} The Figure shows: $a)$ the variation of the per capita income, $b)$ the variation of the properties values, $c)$ the incoming flux of white, and $d)$ the incoming flux of black for the tracts that migrated from one zone to another or to the Overlap zone from 1990 to 2010.}
\label{variationall}
\end{figure}

\begin{figure}[!h]
\centerline{\includegraphics[width=1\textwidth]{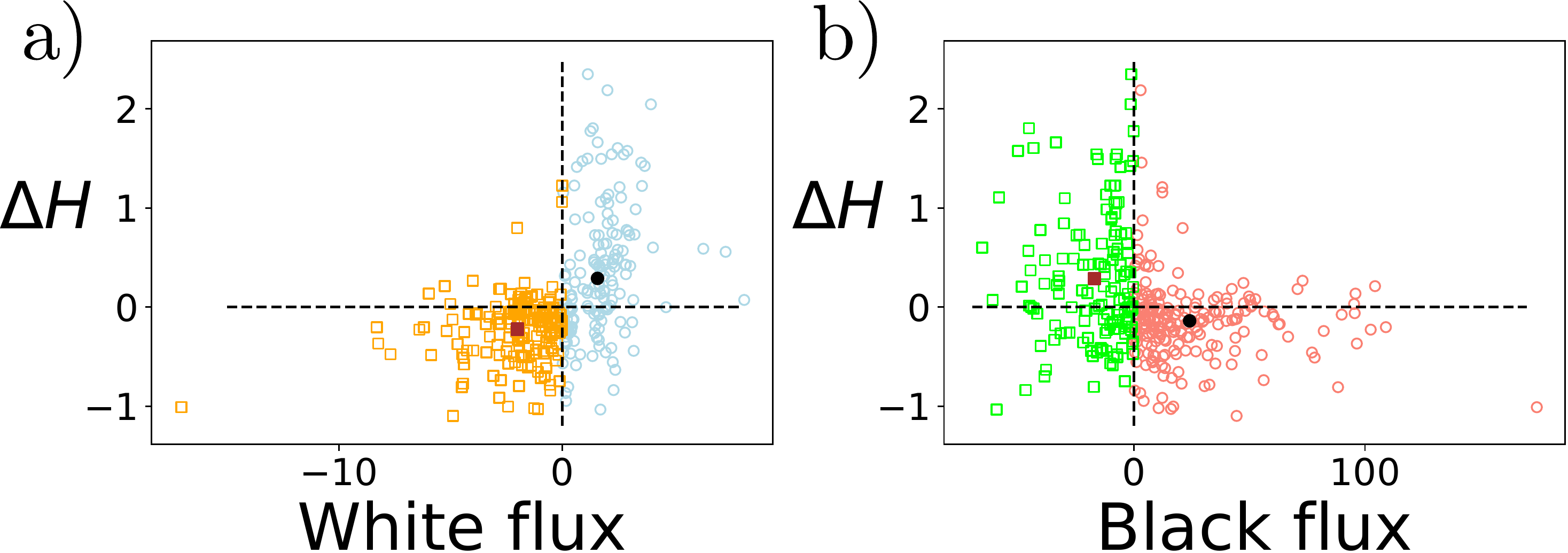}}
\caption{\textbf{Variation of properties values in function of the incoming flux of white and black citizens for the tracts that change zone from 1990 to 2010.} a) Variation of the properties values in function of the incoming flux of white citizens. The tracts with an outgoing flux of white are shown in orange, while the tracts with an incoming flux of white are shown in blue. The black red square is the centroid of the outgoing flux, while the black circle is the centroid of the incoming flux. b) Variation of the properties values in function of the incoming flux of black citizens. The tracts with an outgoing flux of black are shown in green, while the tracts with an incoming flux of black are shown in red. The black red square is the centroid of the outgoing flux, while the black circle is the centroid of the incoming flux. } 
\label{flux_white_and_black}
\end{figure}

In Fig \ref{variationall}, still focusing on the segregation between white and black citizens, we show: the variation of income; the variation of properties values; and the flux of people in the tracts that change zone between the years 1990 and 2010.
For those tracts, in Fig \ref{flux_white_and_black} we compare the variation of the flux of white and black citizens with the variation of the properties values. In Fig \ref{flux_white_and_black}a, we show the outgoing white flux in orange where the red square is the centroid. In blue, we show the incoming white flux, where the black circle is the centroid. While in Fig \ref{flux_white_and_black}b we show the outgoing black flux in green and the red square is the centroid. The incoming black flux in the considered tracts is shown in red and the black circle is the centroid. The figures show that where the flux of white citizens is on average positive, also the properties values increase more than the mean, as well as where the flux of black citizens is negative on average. 

\begin{figure}[!h]
\centerline{\includegraphics[width=1\textwidth]{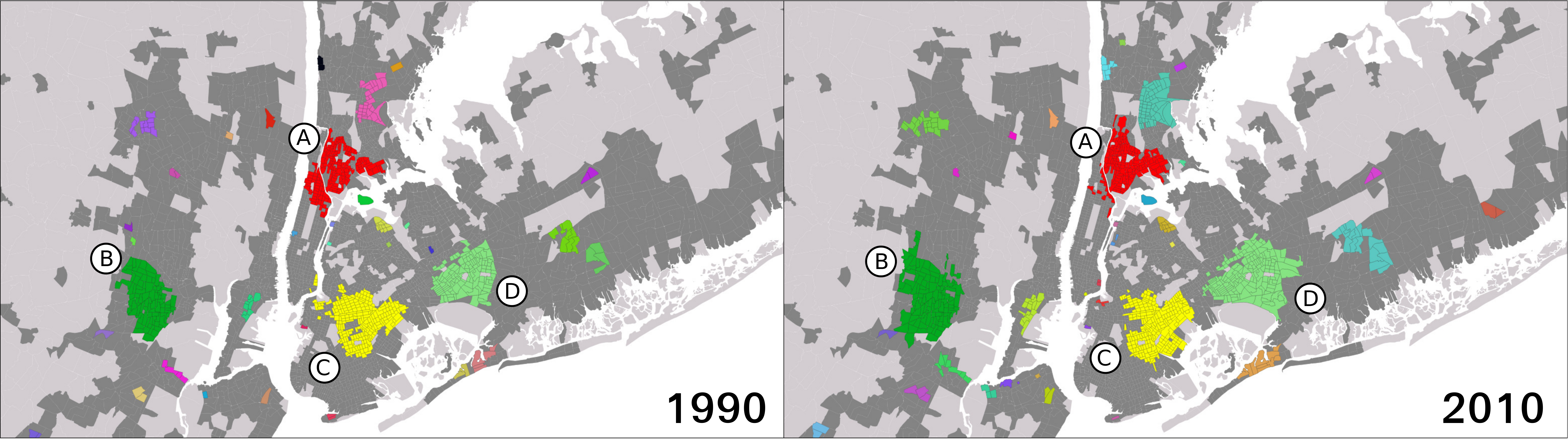}}
\caption{\textbf{Clusterization of the HD black zone for the years of 1990 and 2010.} The Figure shows the results of the clusterization of HD black zone using parameter $\ell' = 1.5~km$ for the years of 1990 and 2010. The four biggest clusters A (in red), B (in dark green), C (in yellow), and D (in light green) are highlighted.} 
\label{dynamics_of_clusters}
\end{figure}

To investigate the dynamics and the displacement of black citizens in New York City, we study the HD black zone. With a simplified version of the CCA we divide in clusters the HD black zone. Indeed, we ignore the threshold $D^*$ and we apply the cutoff length $\ell'$. The parameter $\ell'$ is chosen by analyzing the distribution of the tracts area. Each tract area is considered as a circle with the same area. The mean radius has been found to be $\bar{r}=1.3~km$, therefore in order to consider two neighbors tracts as part of the same cluster, we use $\ell' = 1.5~km$. The results of the clusterization for the years 1990 and 2010 are shown in Fig \ref{dynamics_of_clusters}. In the Figure, we highlight the four biggest clusters A, B, C, and D.

\begin{figure}[!h]
\centerline{\includegraphics[width=0.8\textwidth]{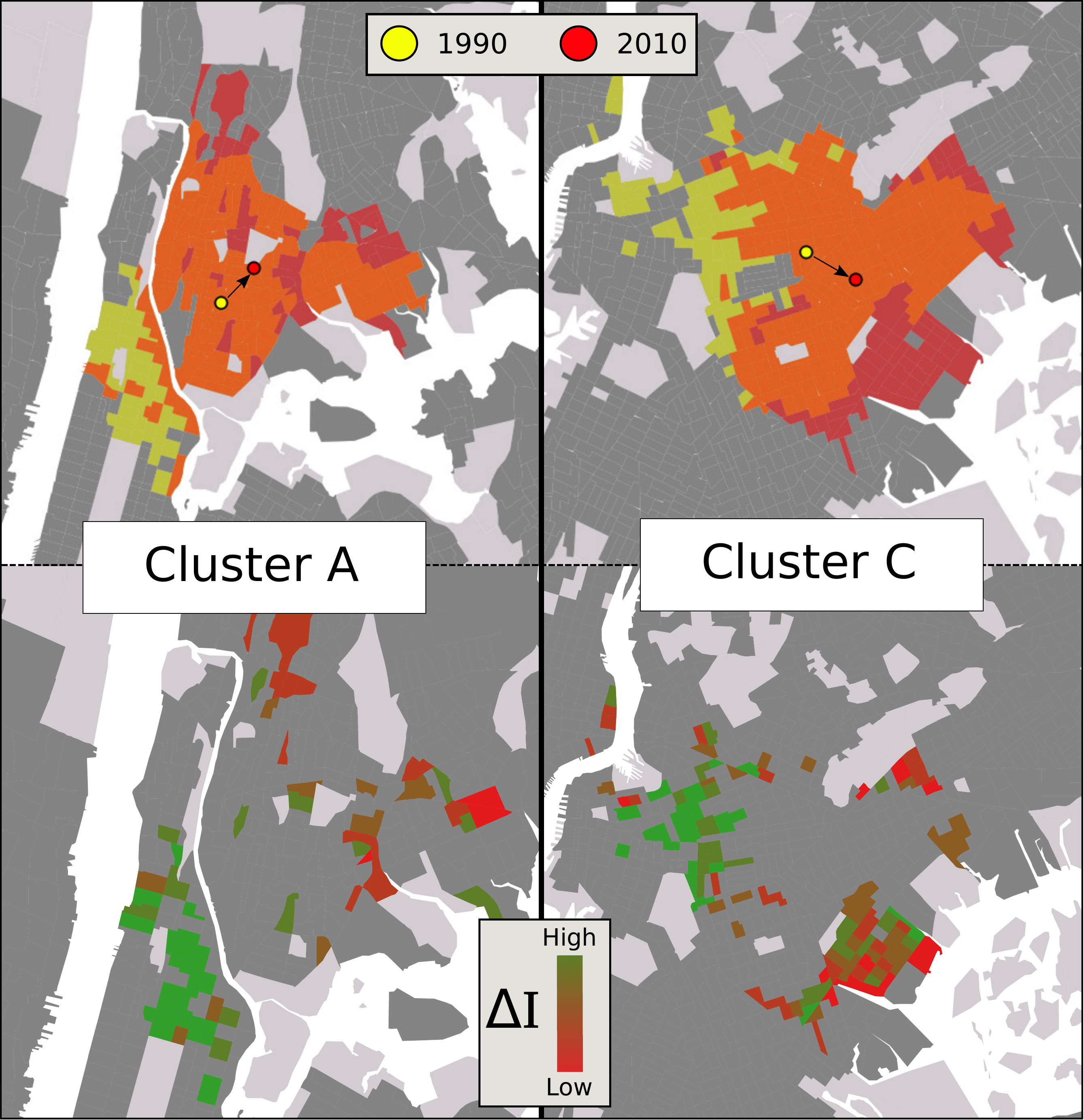}}
\caption{\textbf{Displacement of clusters A and C and the variation of per capita income.} The Figure shows the displacement of cluster A (equivalent to the neighborhood of Harlem and the borough of Bronx) and C (equivalent to the borough of Brooklyn). The clusters in the year 1990 are shown in yellow and the clusters in the year of 2010 are shown in red, with the respective centroids. The figures below show qualitatively the variation of the per capita income for the tracts that change zone in the analyzed period.} 
\label{dynamics_bronx_brooklyin}
\end{figure}

For the four biggest clusters (A, B, C, and D), in Table \ref{displacement} we show the area of each of them for the years 1990 and 2010 and also the displacement of clusters's centroid, highlighting the fact that cluster A and C have a displacement about three times higher than clusters B and D. In Fig \ref{dynamics_bronx_brooklyin}, we show the displacement of clusters A and C from 1990 to 2010. The cluster A includes a region in the neighborhood of Harlem, while the cluster B is inside the boroughs of Brooklyn. In the same Figure, we also show the variation of the per capita income $\Delta I$ for the tracts that change zone in the analyzed period.
\begin{table}[]
\centering
\caption{Areas and displacements of the four biggest clusters of the HD black zone.}
\label{displacement}
\begin{tabular}{l|l|l|l}
           & \textbf{Area$_{1990}$ ($km^2$)} & \textbf{Area$_{2010}$ ($km^2$)} & \textbf{Displacement$_{2010-1990}$ ($km$)} \\ \hline
\textbf{A} &    \multicolumn{1}{c|}{30.7}       &   \multicolumn{1}{c|}{32.8}     & \multicolumn{1}{c}{1.55}   \\ \hline
\textbf{B} &  \multicolumn{1}{c|}{38.0}   &   \multicolumn{1}{c|}{54.6}   &   \multicolumn{1}{c}{0.44}  \\ \hline
\textbf{C} &    \multicolumn{1}{c|}{41.8}       &   \multicolumn{1}{c|}{44.1}     & \multicolumn{1}{c}{1.57}  \\ \hline
\textbf{D} &  \multicolumn{1}{c|}{37.3}   &    \multicolumn{1}{c|}{58.2}   &   \multicolumn{1}{c}{0.64}   \\ 
\end{tabular}
\end{table}

\subsection*{Comparison with the Dissimilarity index}

In order to verify the robustness of our method, we compare the Overlap coefficient defined in Eq \ref{eq_overlap1} with the dissimilarity index \cite{duncan1955}:
\begin{equation}
D_{ab}=\frac{1}{2}\sum_{i=1}^{N}\left |\frac{a_i}{A}-\frac{b_i}{B}  \right |,
\label{eq_dissimilarity}
\end{equation}
where $a_i$ is the population of race $a$ in tract $i$ and $b_i$, the population of race $b$ in the same tract. $A$ and $B$ are the total population of race $a$ and $b$ in the whole city, where the city is defined using the CCA. $N$ are all the tracts that belong to New York City. The value of $D_{ab}$ varies from $0$ to $1$. When it is next to $1$, RRS is high, and vice versa, when it is next to 0 there is not segregation. It shows the percentage of one of the two populations that have to move in order to reduce segregation to $0$ \cite{duncan1955}. The results obtained in New York City are shown in Table \ref{dissimilarityTable}.
\begin{table}[]
\centering
\caption{Dissimilarity index}
\label{dissimilarityTable}
\begin{tabular}{l|l|l|l}
                   & \textbf{1990} & \textbf{2000} & \textbf{2010} \\ \hline
\textbf{White and Black}    & 0.81 & 0.80 & 0.79 \\ \hline
\textbf{White and Hispanic} & 0.64 & 0.64 & 0.62 \\ \hline
\textbf{White and Asian}    & 0.47 & 0.50 & 0.51 \\ \hline
\textbf{Black and Hispanic} & 0.58 & 0.58 & 0.54 \\ \hline
\textbf{Black and Asian}    & 0.78 & 0.78 & 0.76 \\ \hline
\textbf{Hispanic  and Asian} & 0.56 & 0.58 & 0.58 \\ \hline
\end{tabular}
\end{table}

\begin{figure}[!h]
\centerline{\includegraphics[width=0.8\textwidth]{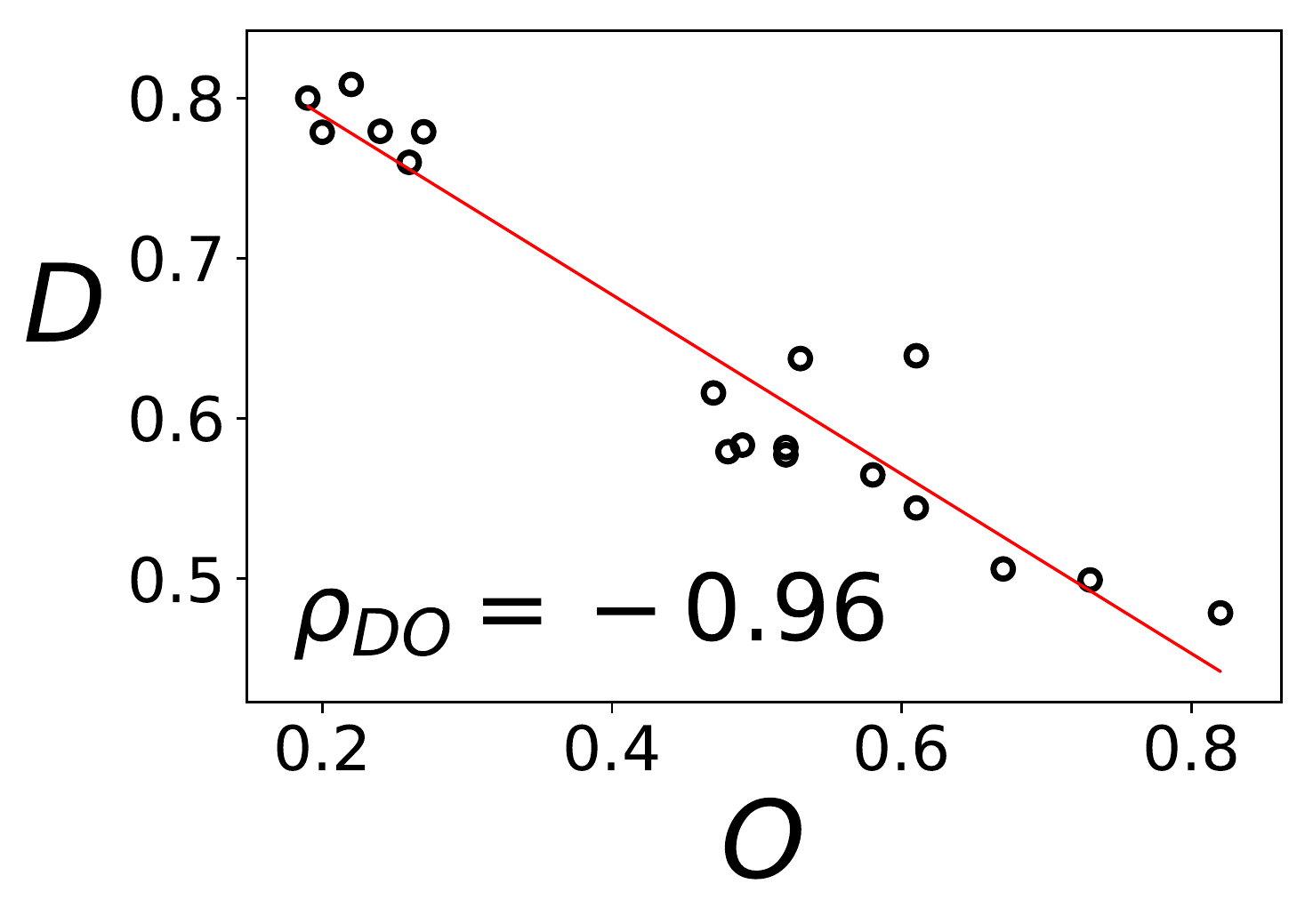}}
\caption{{\bf Dissimilarity index $D$ as a function of the Overlap coefficient $O$.} The red line is the OLS with angular coefficient $m=-0.57\pm 0.1$. The Pearson correlation coefficient, $\rho =-0.96$, shows a strong inverse correlation between the two indexes.}
\label{correlation}
\end{figure}

To analyze the correlation between the two indexes, we plot the dissimilarity indexes $D_{ab}$ found in New York City as a function of their respective Overlap coefficients $O_{rr'}$ (where $X_r$ is the HD zone of race $a$, and $X_{r'}$ of race $b$) in Fig \ref{correlation}. The red line in the Figure shows the result of the Ordinary least Square (OLS). As expected, the relation is inverse with a linear coefficient $m=−0.57\pm 0.01$. Whereupon, in order to quantify the correlation between the two indexes, we calculated the Pearson correlation coefficient (PCC), $\rho_{D,O}=−0.96$. The value implies a strong inverse correlation between the two indexes, proving the robustness of our method.

\section*{Discussion}

We developed a new method in order to measure and to define the topography of RRS and we applied it to the metropolitan area of New York City for the years of  1990, 2000, and 2010. Despite the fact that several studies show that, on average, segregation between white and black citizens in the United States has decreased in the last fifty years \cite{roberto2017,firebaugh2015,logan2010,logan2004}, our results show that it has remained quite stable during the time interval 1990-2010 in the metropolitan area of New York City as well as for black and Asian citizens. Instead, segregation between white and Hispanic, white and Asian, and Hispanic and Asian citizens has grown. Only black and Hispanic are less segregated in 2010 compared with 1990.

By analyzing the per capita income, we observe that white citizens earn more than the other races in all the regions, except when we analyze the segregation between whites and Asian, where Asian citizens have a similar income to white citizens. Regarding the segregation between white and black citizens, we verify that black citizens earn less than white citizens in all the regions. Furthermore, the inequality between white and black citizens is greater in the regions of high density of population of both the races. This result is confirmed by the Gini coefficient, in fact we show that it is higher in the regions of high density of population of two or more races.

Furthermore, we study the segregation between white and black and the segregation between white and Hispanic citizens. We analyze the tracts that change population density from 1990 to 2010 (from region of high density of black, Hispanic, or overlap with white citizens) to region of only high density of white citizens. In this region, we observe that the per capita income and the properties values increased more than the city mean. Conversely, in the tracts that migrated from a region of overlap to a region with high density of population of only black or Hispanic citizens we observe that the per capita income and the properties values increased less than the mean.

Focusing on the segregation between white and black citizens, we analyze the flux of white and black citizens in function of the variation of the properties values. Where the flux of white citizens is positive, the properties values increased more than the city mean, while, where the flux of black citizens is positive, the properties values increased less than the city mean.

Previous studies \cite{zukin2009,lees2003,schaffer1986} questioned the effects of gentrification in the neighborhood of Harlem and in the borough of Brooklyn. Here, by clustering the region of high density of black citizens, we show the displacement of the clusters defined as A (that include a region inside the neighborhood of Harlem) and B (that is inside the borough of Brooklyn). The displacement is of respectively $1.55~km$ and $1.57~km$ in twenty years. This result confirms the theory of displacement of black citizens in the neighborhood of Harlem and in the borough of Brooklyn.

\section*{Acknowledgments}

We gratefully acknowledge CNPq, CAPES, FUNCAP and the National Institute of Science and Technology for Complex Systems in Brazil for the financial support. We especially thank our colleagues and friends of the Complex System group of the Universidade Federal do Cear\'a for the countless discussions. We thank Samuel Morais da Silva and Saulo D. S. Reis for the valuable discussions.

\section*{Appendix A}

\subsection*{Dataset}

All the data used in this paper is extracted from the National Historical Geographic Information System (NHGIS) \cite{manson2017}. The platform provides population, housing, agricultural, and economic data with GIS-compatible boundary files for geographic units in the United States from 1790 to the present. From the platform, population data has been extracted according to race, per capita income data, and the number of owner-occupied housing units by value.

\textbf{Population dataset (TABLE CW7 Persons by Hispanics  or Latino origin by race).} The data provides the number of people for each race for the years of 1990, 2000, and 2010 divided by \textit{Hispanic  or Latino} and \textit{Not Hispanic  or Latino}. We consider white as \textit{Not Hispanic  or Latino: white (single race)}, black as \textit{Not Hispanic  or Latino: black or African American (single race)}, Asian as \textit{Not Hispanic  or Latino: Asian or Pacific Islander (single race)}, and Hispanic  as \textit{Hispanic  or Latino: white (single race)} plus \textit{Hispanic  or Latino: black or African American (single race)} plus \textit{Hispanic  or Latino: Asian   and Pacific Islander (single race)}. The data table is downloadable with the respective GIS-compatible boundary file formed by census tracts standardized to the 2010 census \cite{census}.

\textbf{Per capita income dataset (BD5 Per capita Income in the Previous Year):} The data provides the average per capita income of each American census Tract in the previous year of 1980, 1990, 2000, and between 2008 and 2012. The values are not adjusted for inflation.

\textbf{Properties values dataset (NH23 Specified owner-occupied housing units and B25075 Owner-occupied housing units):} The properties values data are divided into two databases: the table NH23, for the year of 1990, and the table B25075, for the years between 2006 and 2010. The tables provide the number of houses in each price range. The price ranges are divided as: in the table NH23, in twenty ranges, and, in table B25075, in twenty-four ranges from zero Dollar to infinity. For each tract, the weighted arithmetic mean of the properties values has been calculated. The table B25075 is provided in the 2012 census tract and it is consistent with the Population data and the per capita income data, whereas table NH23 is provided in 1990 tracts. Therefore, through a superimposing process, the data was recomposed in the 2012 Census Tract. The superimposing process consists in considering all the properties in a 1990 census tract with centroid in a 2012 census tract as part of that 2012 census tract.

\nolinenumbers


\begin{thebibliography}{10}

\bibitem{chodrow2017} Chodrow P S. Structure and information in spatial segregation. Proceedings of the National Academy of Sciences. 2017. 114 (44): 11591-11597. doi: doi.org/10.1073/pnas.1708201114.

\bibitem{lichter2015} Lichter D T, Parisi D, and Taquino M C. Toward a New Macro-Segregation? Decomposing Segregation within and between Metropolitan Cities and Suburbs. American Sociological Review. 2015,. 80 (4): 843-873. doi: 10.1177/0003122415588558.

\bibitem{fowler2014} Fowler C S. Segregation as a multiscalar phenomenon and its implications for neighborhood-scale research: the case of South Seattle 1990–2010. Urban Geography. 2016. 37 (1): 1-25. doi: dx.doi.org/10.1080/02723638.2015.1043775.

\bibitem{boustan2012} Boustan L P. Racial Residential Segregation in American Cities. The Oxford Handbook of Urban Economics and Planning. 2012. doi: 10.1093/oxfordhb/9780195380620.013.0015.

\bibitem{readon2008a} Readon S F, Farrell C R, Matthews S A, O'Sullivan D, Bischoff K, and Firebaugh G. Race and space in the 1990s: Changes in the geographic scale of racial residential segregation, 1990–2000. Social Science Research. 2008. 38: 55-70. doi: 10.1016/j.ssresearch.2008.10.002.

\bibitem{readon2008b} Readon S F, Mathhews S A, O'Sullivan D, Lee B, Firebaugh G, Farrell C R, and Bischoff K. The geographic scale of metropolitan racial segregation. Demography. 2008. 45 (3): 489-514. doi: doi.org/10.1353/dem.0.0019.  
 
\bibitem{charles2003} Charles C Z. The dynamics of Racial Residential Segregation. Annual Review of Sociology. 2003. 29: 167-207. doi: 10.1146/annurev.soc.29.010202.100002. 

\bibitem{massey1993} Massey D S and Denton N A. American Apartheid. Harvard Univ Pr. isbn-10: 0674018214.

\bibitem{roberto2017} Roberto E, and Hwang J. Barriers to Integration: Physical Boundaries and the Spatial Structure of Residential Segregation. arXiv. 2017. arXiv: 1509.02574.

\bibitem{firebaugh2015} Firebaugh G, and Farell C R. Still Large, but Narrowing: The Sizable Decline in Racial Neighborhood Inequality in Metropolitan America, 1980–2010. Demography. 2016. 53 (1): 139-64. doi: 10.1007/s13524-015-0447-5.

\bibitem{logan2010} Logan J R,  and Stults B J. The Persistence of Segregation in the Metropolis: New Findings from the 2010 Census. 2010. Accessed 9th of March 2018. url: \url{https://s4.ad.brown.edu/Projects/Diversity/Data/Report/report2.pdf}

\bibitem{logan2004} Logan J R, Stults B J, and Farley R. Segregation of minorities in the metropolis: Two decades of change. Demography. 2004. 41 (1): 1-22.  

\bibitem{roberto2016} Roberto E. The Divergence Index: A Decomposable Measure of Segregation and Inequality. arXiv. 2015. arXiv: arXiv:1508.01167.

\bibitem{louf2016} Louf R, and Barthelemy M. Patterns of residential segregation. PLoS ONE. 2016. 11 (6): e0157476. doi: doi.org/10.1371/journal.pone.0157476.

\bibitem{roberto2015} Roberto E. The Spatial Proximity and Connectivity (SPC) Method for Measuring and Analyzing Residential Segregation. arXiv. 2016. doi: arXiv:1509.03678. 

\bibitem{jargowsky2005} Jargowsky P A, Kim J,  A Measure of Spatial Segregation: The Generalized Neighborhood Sorting Index. National Poverty Center Working Paper Series. Accessed the 12th March 2018. url: \url{http://www.npc.umich.edu/publications/working_papers/}.

\bibitem{readon2004} Readon S F, and O'Sullivan D. Measures of Spatial Segregation. Sociological Methodology. 2004. 34: 121-162. 

\bibitem{readon2002} Readon S F, and Firebaugh G. Measures of Multigroup Segregation. Sociological Methodology. 2002. 32 (1): 33-67. doi: doi.org/10.1111/1467-9531.00110.

\bibitem{massey1988} Massey D, and Denton N. The Dimensions of Residential Segregation. Social Forces. 1988. 67 (2): 281-315. doi: 10.2307/2579183.   

\bibitem{winship1977} Winship C. A Revaluation of Indexes of Residential Segregation. Social Forces. 1977. 55 (4): 1058-1066. doi: 10.2307/2577572.

\bibitem{duncan1955} Duncan O D, and Duncan B. A Methodological Analysis of Segregation Indexes. American Sociology Rev. 1955. 20: 210-217. doi: 10.2307/2088328.

\bibitem{watson2009} Watson T. Inequality and the measurement of of residential segregation by income in American neighborhoods. The review of income and wealth. 2009. 55 (3): 820-844. doi: 10.1111/j.1475-4991.2009.00346.x.

\bibitem{acevedo-garcia2003} Acevedo-Garcia D, Lochner K A, Osypuk T L, and Subramanian S V. Future Directions in Residential Segregation and Health Research: A Multilevel Approach. The American Journal of Public Health. 2003. 93 (2): 215-221. 

\bibitem{williams2001} Williams D R, and Collins C. Racial Residential Segregation: A Fundamental Cause of Racial Disparities in Health. Public Health Reports. 2001. 116 (5): 404-416. doi: doi.org/10.1093/phr/116.5.404.

\bibitem{fang1998} Fang J, Madhavan S, Bosworth W, and Alderman M H. Residential segregation and mortality in New York City. 1998. 47 (4): 469-76. doi: doi.org/10.1016/S0277-9536(98)00128-2.

\bibitem{helbrecht2018} Helbrecht I. Gentrification and Displacement. SpringerVS. doi: doi.org/10.1007/978-3-658-20388-7\_1.

\bibitem{brown2017} Brown-Saracino J. Explicating Divided Approaches to Gentrification and Growing Income Inequality. Annual Review of Sociology. 2017. 43: 515-39. doi: doi.org/10.1146/annurev-soc-060116-053427.

\bibitem{lees2016} Lees L. Gentrification, Race, and Ethnicity: Towards a Global Research Agenda? City and Community. 2016. 208-214. doi: 10.1111/cico.12185.

\bibitem{freeman2008} Freeman L. Neighbourhood Diversity, Metropolitan Segregation and Gentrification: What Are the Links in the US? UrbanStudies. 2008. 46 (10): 2079-2101. doi: 10.1177/0042098009339426.

\bibitem{freeman2004} Freeman L, and Braconi F. Gentrification and Displacement New York City in the 1990s. Journal of the American Planning Association. 2007. 39-52. doi: doi.org/10.1080/01944360408976337.

\bibitem{smith1998} Smith N. The Encyclopedia of Housing. London: Taylor \& Francis. 1998

\bibitem{curran2006} Curran W. ``From the Frying Pan to the Oven'': Gentrification and the Experience of Industrial Displacement in Williamsburg, Brooklyn. 2007. 44 (8): 1427-1440. doi:  10.1080/00420980701373438.

\bibitem{zuk2017} Zuk M, Bierbaum A H, Chapple K, Gorska K, and Loukaitou-Siders A. Gentrification, Displacement, and the Role of Public Investment. Journal of Planning Literature. 2017. 33 (1): 31-44. doi: 10.1177/0885412217716439.

\bibitem{zukin2009} Zukin S. New Retail Capital and Neighborhood Change: Boutiques and Gentrification in New York City. City and Community. 2009. doi: 10.1111/j.1540-6040.2009.01269.x.

\bibitem{lees2003} Lees L. Super-gentrification: The Case of Brooklyn Heights, New York City. Urban Studies. 2003. 40 (13): 2487-2509. doi: 10.1080/0042098032000136174.

\bibitem{schaffer1986} Schaffer R, and Smith N. The Gentrification of Harlem? Annals of the Association of American Geographers banner. 1986. 76 (3): 347-365. doi: 10.1111/j.1467-8306.1986.tb00124.x.

\bibitem{census} US Census Bureau. Accessed 9th of March 2018. url: \url{ttp://www.census.gov}

\bibitem{operti2017} Operti F G, Oliveira E A, Carmona H A, Machado J C, and Andrade J S. The light pollution as a surrogate for urban population of the US cities. Physica A. 2018. 492: 1088–1096. doi: https://doi.org/10.1016/j.physa.2017.11.039.

\bibitem{caminha2017} Caminha C, Furtado V, Pequeno T H C, Ponte C, Melo H P M, Oliveira E A, Andrade J S. Human mobility in large cities as a proxy for crime. PLoS One. 2017. 12: 2 e0171609. doi:dx.doi.org/10.1371/journal.pone.0171609.

\bibitem{oliveira2014} E.A. Oliveira, J.S. Andrade, and H.A. Makse. Large cities are less green. Scientific Report. 2015. 4: 4235. doi: http://dx.doi.org/10.1038/srep04235.

\bibitem{rozenfeld2011} Rozenfeld H D, Rybski D, Gabaix X, Makse H A. The area and population of cities: New insights from a different perspective on citie. American Economic Review. 2011. 101 (5): 2205. doi:dx.doi.org/10.1257/aer.101.5.2205.

\bibitem{rozenfeld2008} Rozenfeld H D, Rybski D, Andrade J S, Batty M, Stanley H E, and Makse H A. Laws of population growth. Proceedings of the National Academy of Sciences. 2008. 105 (48): 18702-18707. doi:dx.doi.org/10.1073/pnas.0807435105.

\bibitem{bettencourt2007} Bettencourt L M A, Lobo J, Helbing D, K\"uhnert C, and West G B. Growth, innovation, scaling, and the pace of life in cities. Proceedings of the National Academy of Sciences. 2007. 104: 7301-7306. doi: doi.org/10.1073/pnas.0610172104.

\bibitem{makse1998} Makse H A, Andrade J S, Batty M, Havlin S, Stanley H E. Modeling urban growth patterns with correlated percolation. Physical Review E. 1998. 58: 7054–7062. doi: dx.doi.org/10.1103/PhysRevE.58.7054.

\bibitem{makse1995}  Makse H A, Havlin S, Stanley H E. Modelling urban growth patterns. Nature. 1995. 377: 6550 608. doi:dx.doi.org/10.1038/377608a0.

\bibitem{arcaute2015} Arcaute E, Ferguson P, Youn H, Johansson A, and Batty M. Constructing cities, deconstructing scaling laws. Royal Society. 2015. 12: 102. doi:doi.org/10.1098/rsif.2014.0745.

\bibitem{arcaute2016} Arcaute E, Molinero C, Hatna H, Murcio R, Vargas-Ruiz C, Masucci A P, and Batty M. Cities and regions in Britain through hierarchical percolation. 2016. 3: 4. doi:doi.org/10.1098/rsos.150691.

\bibitem{vijaymeena2016} Vijaymeena M K and Kavitha K. A survey on similarity measures in text mining. Machine Learning and Applications: An International Journal (MLAIJ). 2016. 3 (1). doi:10.5121/mlaij.2016.3103. 

\bibitem{gini1912} Gini C. Variabilità e mutabilità. Reprinted in Pizetti. Salvemini. Memorie di metodologica statistica. 1955. Rome: Libreria Eredi Virgilio Veschi.

\bibitem{manson2017} Manson S, Schroeder J, Riper D V, and Ruggles S. IPUMS National Historical Geographic Information System: Version 12.0 [Database]. Minneapolis: University of Minnesota. 2017. doi: doi.org/10.18128/D050.V12.0.

\end{thebibliography}
\end{document}